\documentclass[twocolumn,twoside]{Jornadas}
\usepackage[utf8]{inputenc}
\usepackage[spanish]{babel}
\usepackage{listings}
\usepackage{algorithm}
\usepackage{algorithmicx}
\usepackage{algcompatible}
\usepackage{adjustbox}
\usepackage{graphicx}
\usepackage{color}
\usepackage{caption}
\usepackage[caption=false,font=footnotesize]{subfig}
\usepackage{placeins}
\usepackage{hyperref}
\usepackage{url}

\definecolor{gray97}{gray}{.97}
\definecolor{gray75}{gray}{.75}
\definecolor{gray45}{gray}{.45}

\def\BibTeX{{\rm B\kern-.05em{\sc i\kern-.025em b}\kern-.08em
    T\kern-.1667em\lower.7ex\hbox{E}\kern-.125emX}}

\graphicspath{{.}{./Figuras/}}

\begin{document}

\title{Técnicas Quantum-Inspired en Tensor Networks para Contextos Industriales}

\author{%
     Alejandro Mata Ali%
     \thanks{i3B, Quantum Development Department, Paseo Mikeletegi 5, 20009 Donostia, España, e-mail: {\tt amata@ayesa.com}},
     Iñigo Perez Delgado
     \thanks{i3B, Parque Tecnológico de Bizkaia, Ibaizabal Bidea, Edif. 501-A, 48160 Derio, España, e-mail: {\tt iperezde@ayesa.com}}
     y Aitor Moreno Fdez. de Leceta%
     \thanks{i3B, Unidad de Inteligencia Artificial, Avenida de los Huetos, Edificio Azucarera, 01010 Vitoria, España, 
     e-mail: {\tt aimorenof@ayesa.com}}
}

\maketitle
\markboth{}{}
\pagestyle{empty} 
\thispagestyle{empty} 

\begin{abstract}
En este artículo presentamos un estudio de la aplicabilidad y viabilidad de los algoritmos y técnicas de inspiración cuántica en tensor networks para entornos y contextos industriales, con una recopilación de la literatura disponible y un análisis de los casos de uso que pueden verse afectados por dichos métodos. Además, exploramos las limitaciones de dichas técnicas a fin de determinar su posible escalabilidad.
\end{abstract}



\section{Introducción}
\PARstart{L}{a} computación cuántica es un campo novedoso que está recibiendo un notable interés debido a su particular manera de realizar operaciones y procesar información, siendo capaz de abordar problemas computacionales altamente complejos. En la era de la información y la digitalización en la que nos encontramos, esta tecnología tiene una notable aplicabilidad para casos industriales, por ejemplo en la optimización de rutas de reparto \cite{TSP_Quantum1,TSP_Quantum2}, el almacenamiento de paquetes \cite{Bin_Quantum} o el aprendizaje automático (o `machine learning') cuántico \cite{QML} aplicado a diversos contextos \cite{Vision,Quanvolutional,Class1,Class2}.

Sin embargo, a pesar de los grandes avances en el desarrollo de los ordenadores cuánticos digitales (o `de puertas lógicas'), persisten desafíos sustanciales como la fragilidad de los estados cuánticos, los errores en las operaciones cuánticas o la escalabilidad del número de qubits.

En este contexto, surge la necesidad de explorar enfoques alternativos que puedan aprovechar las propiedades de cálculo de los sistemas cuánticos, pero que no tengan que ejecutarse en los ordenadores cuánticos digitales. Uno de estos enfoques puede ser el recocido cuántico (o `quantum annealing') \cite{Annealing,Dwave}, que se ejecuta en dispositivos cuánticos especializados dedicados a la optimización combinatoria de funciones de coste concretas. Otro es el recocido digital (o `digital annealing') \cite{Digital,Digital2,Digital3}, que se inspira en propiedades del Quantum Annealing para simularlo clásicamente.

Por otro lado, una de las tecnologías más prometedoras, aunque más desconocida dentro de la industria, son las tensor networks \cite{TN}. Las tensor networks son una clase de algoritmos y técnicas de inspiración cuántica basadas en imitar las operaciones tensoriales que realiza un ordenador cuántico, pero ejecutándolas en ordenadores clásicos. Mediante el uso de propiedades tensoriales se puede optimizar la ejecución de dichas operaciones, sobre todo en casos en los cuales no se requiere todo el vector de estado cuántico, sino solo propiedades del mismo.

Otra característica relevante de las tensor networks es que permiten representar de manera eficiente determinadas familias de estados cuánticos, mediante representación como el matrix product state (MPS)\cite{MPS1,MPS2}, también llamadas tensor train (TT), o projected entangled pair states (PEPS) \cite{MPS2}. De esta manera, con una cantidad reducida de memoria podemos realizar cálculos cuánticos y obtener propiedades de sistemas complejos. Estas mismas representaciones han sido relevantes para el mundo del machine learning, ya que son capaces de comprimir modelos reduciendo la memoria requerida sin perder una cantidad notable de precisión \cite{Compress1,Compress2,ML_TN}.

Dadas todas estas capacidades, las tensor networks son altamente susceptibles de ser utilizadas en contextos industriales, pudiendo abordar problemas altamente complejos y de gran tamaño de manera eficiente. A pesar de ser un campo relativamente maduro a nivel académico (iniciado en 1971 por Penrose \cite{Penrose}), su estudio aplicado a la industria es reciente y siempre asociado a la computación cuántica. Sin embargo, dados los grandes resultados arrojados en los últimos años por las tensor networks, superando en desempeño a los algoritmos cuánticos ejecutables en el hardware actual \cite{Q_Adv1,Q_Adv2}, muchas empresas están empezando a interesarse en el estudio e implementación de esta tecnología.

En este trabajo hacemos un estudio de los principales usos de las tensor networks para casos de uso industriales, haciendo una recopilación de artículos del estado del arte y exponiendo los puntos fuertes y debilidades de cada técnica. Además, clasificamos los algoritmos según su caso de uso específico, escala a la que se quiera aplicar y desempeño requerido. El campo de las tensor networks dispone de una literatura extensa y compleja, por lo que el objetivo de este artículo es servir como una guía introductoria al mismo.

\section{Introducción a las tensor networks}
Lo primero que haremos será realizar una breve introducción a las tensor networks a fin de que se entiendan las seciones posteriores. Con el propósito de no salirnos del objetivo principal del artículo, asumiremos conocimiemto de conceptos básicos de álgebra lineal y nos centraremos en las tensor networks como tal. Existe información complementaria sobre tensor networks en \cite{TN,TN2}.

\begin{figure}[htb] 
\begin{center} 
  \includegraphics[width=7.5cm]{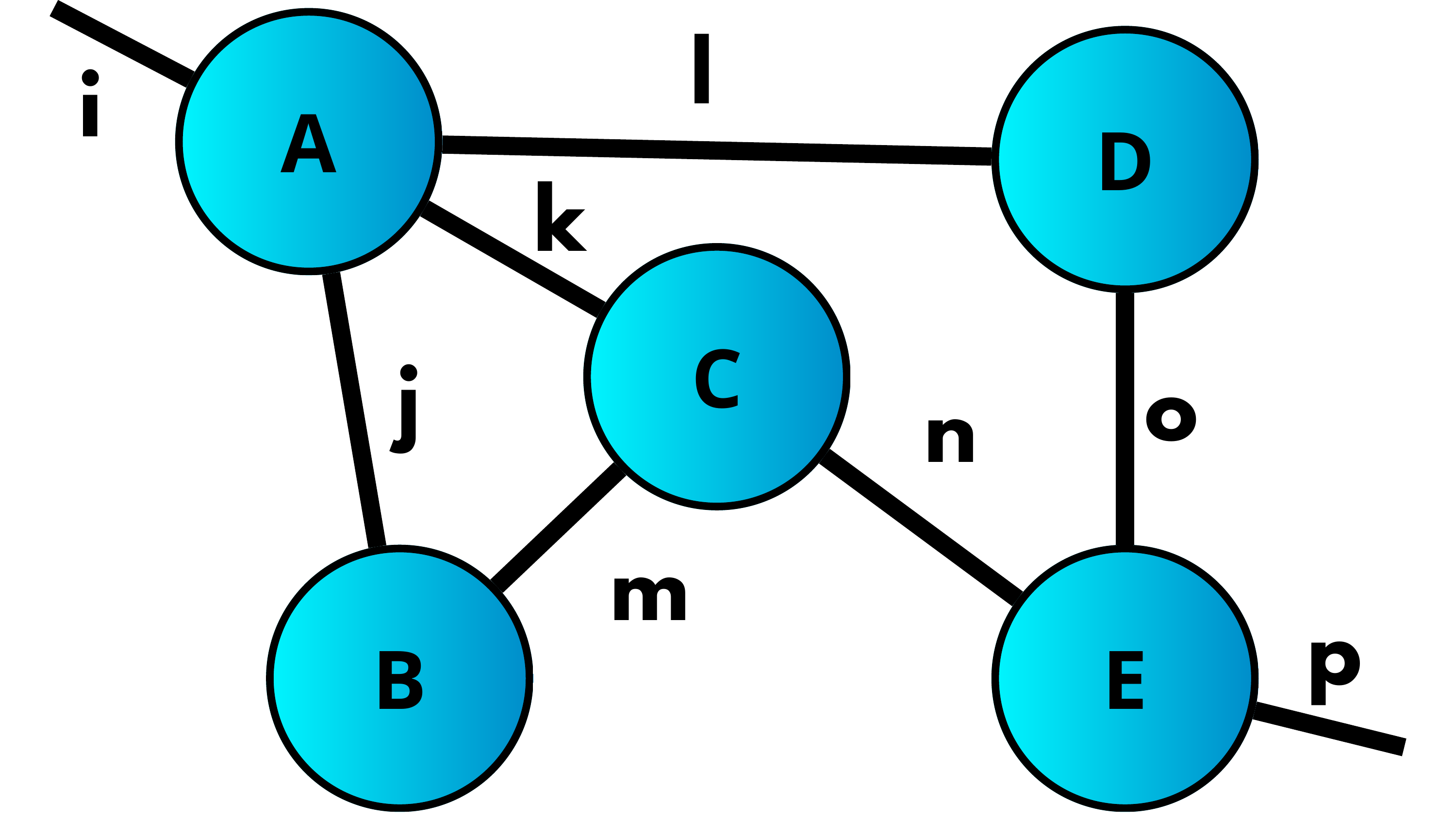}
\end{center} 
\caption{Tensor network representando el 2-tensor $T_{ip}$ de la Ec. \ref{eq: TN General}.} 
\label{fig: TN general} 
\end{figure}

Una tensor network (TN) es una representación gráfica de una ecuación multilineal entre diferentes tensores. Podemos ver un ejemplo en la Fig. \ref{fig: TN general}, donde tenemos la tensor network que representa el 2-tensor $T_{ip}$ cuyos elementos se obtienen mediante la operación de contracción
\begin{equation}\label{eq: TN General}
    T_{ip} = \sum_{j,k,l,m,n,o} A_{ijkl}B_{jm}C_{kmn}D_{lo}E_{nop}.
\end{equation}

Como se puede ver, esta es una representación muy simple de una operación muy compleja, en cuyas generalizaciones más grandes resulta complicado determinar las relaciones entre los tensores que se están contrayendo y la manera más eficiente de contraerlos. Además, podemos abstraer más la tensor network eliminando los nombres de los índices y los tensores si sabemos por alguna regla cuáles deberían ser. Un ejemplo sería que los índices del tensor se vayan colocando según la orientación de las agujas del reloj.

Las reglas que siguen dichas representaciones son las siguientes:
\begin{enumerate}
    \item Cada nodo del grafo representa un tensor.
    \item Cada línea saliente de un tensor representa un índice de dicho tensor. Por tanto, un nodo con $n$ líneas salientes representa un $n$-tensor.
    \item Cada línea que conecta dos tensores representa que ambos tensores están contraidos con respecto a un índice mudo común.
    \item Las líneas que solo salgan de un tensor, pero no conecten con otro, representan índices libres. Por tanto, una tensor network con $n$ índices libres representa un $n$-tensor.
    \item Los nodos que no estén conectados por un índice quedan en producto tensorial entre ellos.
\end{enumerate}

\begin{figure}[htb] 
\begin{center} 
  \includegraphics[width=7.5cm]{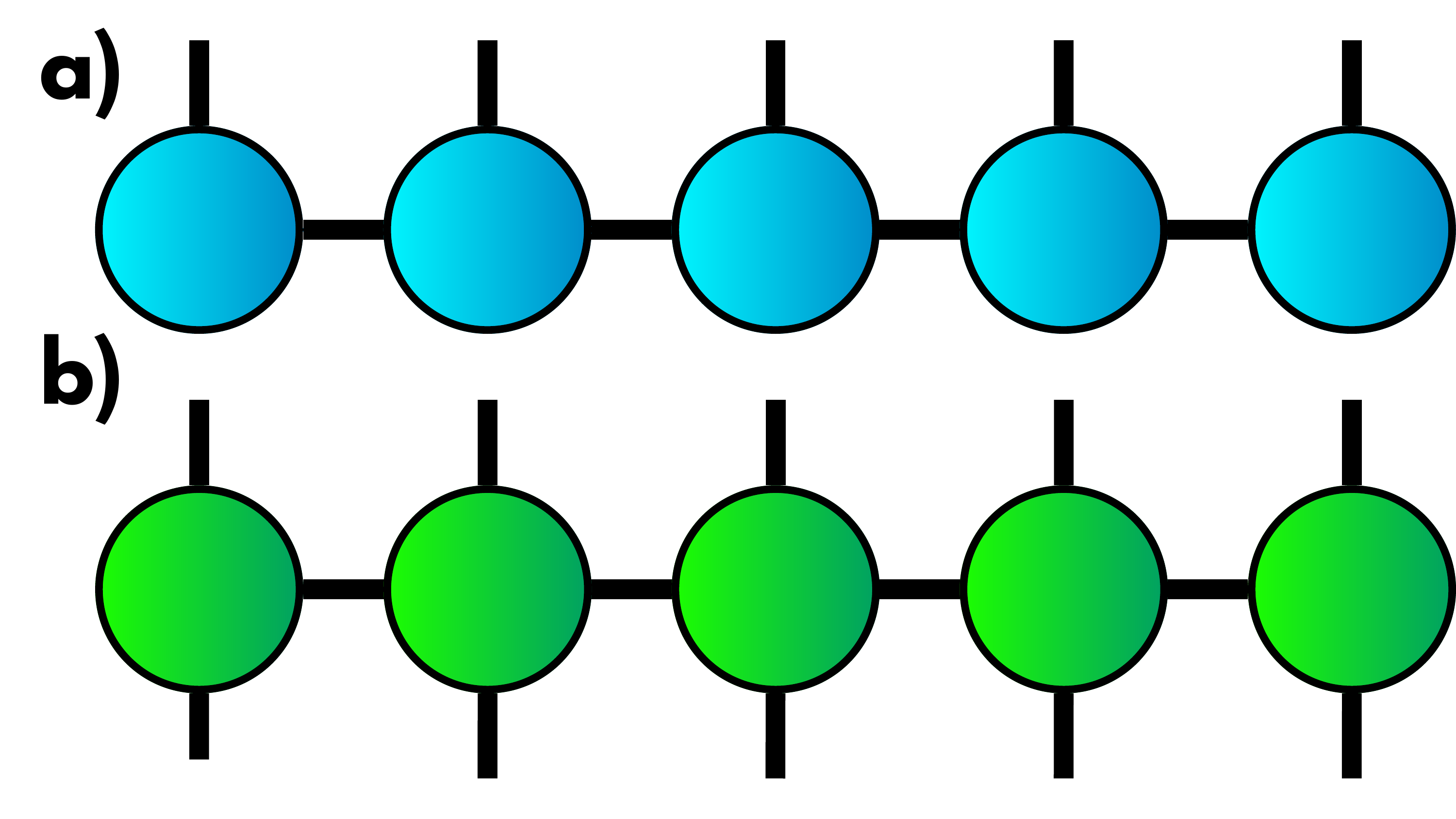}
\end{center} 
\caption{a) Forma MPS de un 5-tensor, b) Forma MPO de un 10-tensor.} 
\label{fig: MPS y MPO} 
\end{figure}

Un tipo altamente interesante y útil de tensor network es el ya mencionado MPS\cite{MPS1}, ya que consiste en una representación unidimensional y descompuesta de un tensor mediante el producto de matrices. Podemos ver un ejemplo en la Fig. \ref{fig: MPS y MPO} a), donde vemos la representación en forma MPS de un 5-tensor. Esta representación MPS está compuesta de 2 tipos de índices: los índices de enlace, que tendrán una dimensión $b$, y los índices físicos, de dimensión $d$, correspondientes a las dimensiones del tensor a representar. Los índices físicos son los índices libres, que coinciden con los índices del tensor a representar, mientras que los índices de enlace son los que conectan a los tensores de la red entre sí y se encargan de dar cuenta de las correlaciones entre los índices físicos. Hay que destacar que existen otras versiones de la representación MPS con diferentes condiciones de frontera (la manera de enlazar los nodos de los extremos), pero la más simple es esta. También existe la posibilidad de que las dimensiones de los índices de enlace sean diferentes, al igual que las de los índices físicos, por cuestiones que mencionaremos al final de esta sección.

Para una representación MPS $N$-tensorial, con dimensión de enlace constante $b$ y dimensión física constante $d$, el número de elementos que hay que almacenar es $2db + (N-2)db^2$ frente a los $d^N$ que haría falta en la versión densa, contraida, del tensor, $db$ por cada nodo de los extremos y $db^2$ por cada nodo del interior de la cadena.

Todo tensor puede representarse en representación MPS de manera exacta, aunque no siempre eficientemente, ya que $b$ puede ser del orden de $d^N$ en determinados casos complejos y en los que la correlación es muy fuerte. Por tanto, será importante considerar cómo aumenta $b$ con el tamaño del tensor a representar. Aun así, muchos casos de interés pueden representarse en representación MPS de manera eficiente. Por ejemplo, en contextos académicos su uso más extendido es el de la de representar estados cuánticos de $N$ partículas, esto es, vectores de estado de $d^N$ componentes.

Una extensión de la representación MPS es el Matrix Product Operator (MPO) \cite{MPO}, consistente en permitir que cada nodo disponga de 2 índices físicos, como vemos en la Fig. \ref{fig: MPS y MPO} b). La utilidad de esta representación reside en los casos en los que queramos representar una matriz de $d^{2N}$ componentes, como una interacción cuántica. En este caso, el número de elementos a almacenar es de $2d^2b+(N-2)d^2b^2$.

Dichas formas pueden obtenerse principalmente de tres maneras. La primera es mediante la construcción lógica de la representación, de forma que las interacciones entre los diferentes nodos de la tensor network generen el tensor que deseamos. La segunda será la construcción variacional, consistente en empezar con un conjunto de tensores prueba y utilizar técnicas como el descenso de gradiente para conseguir que se convierta en el que queremos representar. La tercera es mediante la descomponsición de valor singular, o `singular value decomposition' (SVD) \cite{SVD}, iterativa, donde realizamos una SVD por cada par de índices físicos.

Es importante destacar que se puede usar cualquiera de las dos representaciones para representar un vector, matriz, o tensor general, ya que podemos convertir un $m$-tensor en un $n$-tensor mediante técnicas como el agrupamiento y la separación de índices (`grouping' y `splitting' en inglés) \cite{TN}, un mapeado biyectivo de un conjunto de índices a otro conjunto de índices.

Pongamos un ejemplo obteniendo la representación MPO de $n$ nodos de una matriz $N\times M$. Para ello tendremos que empezar transformando la matriz en un $2n$-tensor, mediante un splitting. El tensor $A'$ resultante tendrá unas dimensiones $\vec{D}$ tales que
\begin{equation}
    N = \prod_{i=0}^{n-1} D_i,\qquad M = \prod_{i=n}^{2n-1} D_i.
\end{equation}

\begin{figure}[htb] 
\begin{center} 
  \includegraphics[width=7.5cm]{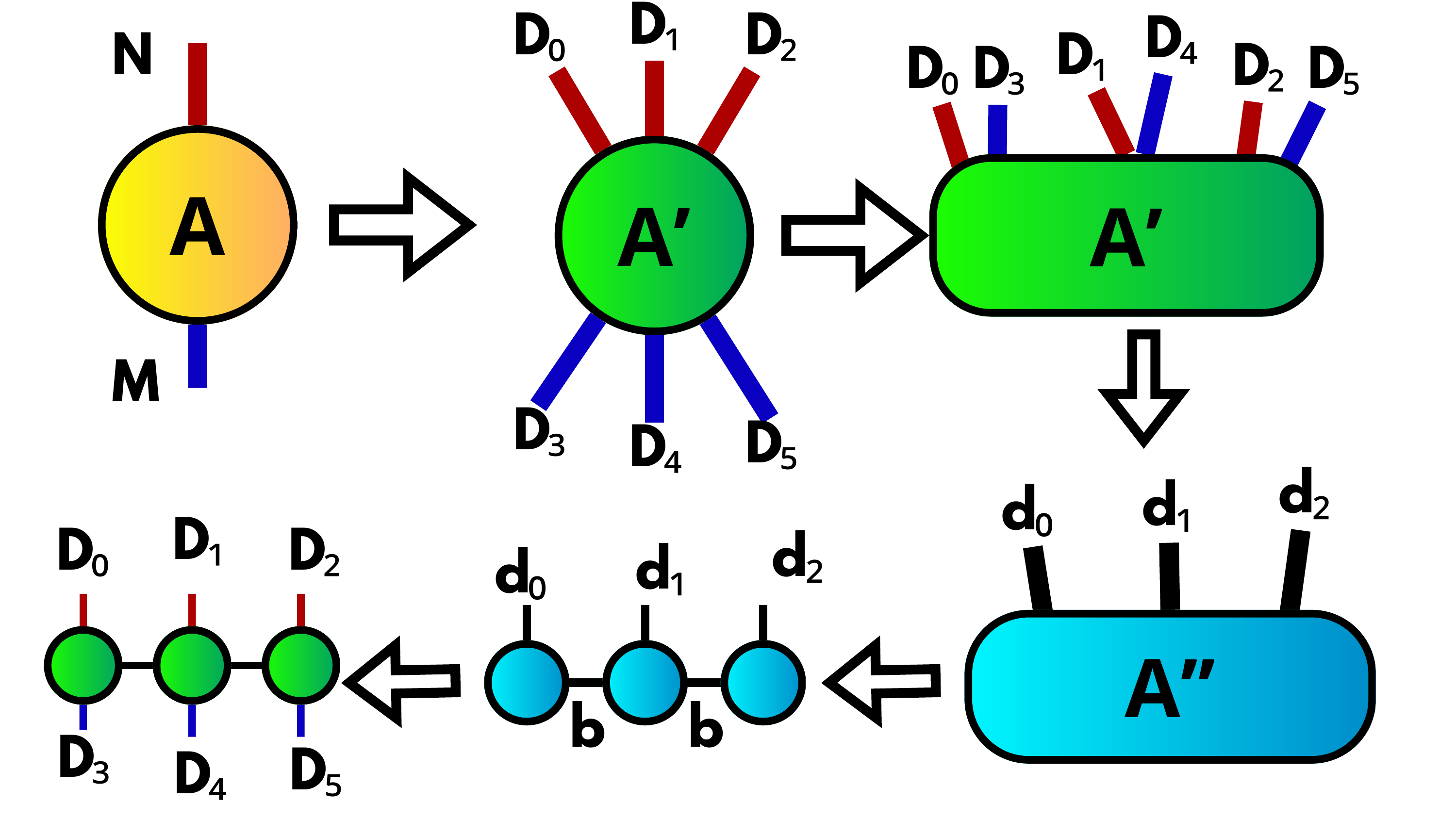}
\end{center} 
\caption{Proceso de compresión de una matriz a su representación MPO. 1) Aplicamos un splitting. 2) Juntamos lo índices de cada nodo por parejas. 3) Aplicamos un grouping de las parejas de índices. 4) Realizamos la SVD iterativa. 5) Aplicamos un splitting a cada índice físico.} 
\label{fig: Compresion_Capa} 
\end{figure}

Ahora juntaremos las parejas de índices entrada y salida que queremos que sean parte del mismo nodo de la forma MPO y hacemos grouping entre ellos para obtener un tensor $A''$ cuyas nuevas dimensiones $\vec{d}$ son $d_i=D_{2i}D_{2i+1}$. Ahora realizamos el proceso iterativo de búsqueda de la representación MPS de ese tensor, para finalmente hacer un splitting de los índices físicos de cada nodo para obtener así los índices de entrada y salida originales. Todo el proceso lo podemos ver en la Fig.\ref{fig: Compresion_Capa}.

Si en vez de desear una representación exacta de un tensor, quisiéramos una representación aproximada del mismo con el fin de tener un menor $b$ y, por tanto, ahorrar memoria, podremos aprovechar estas dos representaciones. En la construcción variacional simplemente operaremos con un menor $b$, mientras que en la SVD iterativa realizaremos SVD truncadas (TSVD) \cite{TSVD}, quedándonos solo con los $b$ mayores valores singulares, que representan las correlaciones más significativas. La precisión de esta aproximación dependerá tanto del tensor en sí que queramos representar como del mapeo que usemos, ya que si juntamos en los mismos nodos las `columnas' más correlacionadas, todas sus correlaciones estarán dentro del propio nodo. Además, si tenemos de forma adyacente los índices más correlacionados entre sí, necesitaremos una menor dimensión de enlace, ya que la intensidad de las correlaciones entre dos nodos de una representación MPS decae exponencialmente con la distancia geodésica entre ellos, la distancia mínima que los separa.

\section{Propiedades tensoriales básicas y técnicas más complejas}
En esta sección vamos a realizar una conexión entre las propiedades básicas de las tensor networks y ciertas técnicas más avanzadas y complejas. De esta manera crearemos un puente entre la abstracción de las tensor networks y sus aplicaciones prácticas. Para ello, abordaremos cada caso explicando qué propiedades son determinantes en el mismo y cuál es la ventaja que ofrece con respecto a otros métodos.

\subsection{Simulación de sistemas cuánticos}
Como es evidente por su contexto histórico, las tensor networks están especialmente pensadas para la representación eficiente de sistemas físicos de muchos cuerpos. Esto es de gran interés en campos como la física de la materia condensada \cite{Many1,Many2} o la computación cuántica \cite{Noisy_QC,TN_QC_Circuit}.

Para ello se escoge una solución estimada (o `ansatz') que sea representable como una tensor network y se trabaja con ella, traduciendo a notación tensorial también operadores físicos y magnitudes. Casos en los que hay un gran éxito por parte de las tensor networks es en la obtención de estados de mínima energía en química cuántica \cite{Q_Chemistry} y en la evolución temporal de ciertos estados cuánticos para obtener ciertas magnitudes y propiedades de los mismos \cite{Q_Adv1,Q_Adv2}.

\subsection{Compresión de modelos de machine learning}\label{ssec: mach}
La más conocida aplicación de las tensor networks es la compresión de modelos de machine learning \cite{Compress1,Compress2,ML_TN}. Esto se puede realizar tanto comprimiendo un modelo ya creado y entrenando o creando un modelo comprimido y entrenar la tensor network que lo representa directamente. La clave de la compresión de modelos reside en encontrar una representación eficiente del modelo que conserve sus propiedades de análisis. Vamos a explicar cómo funciona mediante la compresión de una capa densa en una red neuronal a sus representaciones MPS y MPO aproximadas.

Para empezar, tendremos que tener en cuenta que la acción de una capa densa de una red neuronal, antes de la acción de una función de activación, se puede escribir como
\begin{equation}
    \vec{v} = A\vec{x} +\vec{c},
\end{equation}
donde $A$ es una matriz de dimensiones $N\times M$, $\vec{v}$ y $\vec{c}$ son vectores de dimensión $N$ y $\vec{x}$ es un vector de dimensión $M$. La entrada a la capa es el vector $\vec{x}$, mientras que la salida será el vector $\vec{v}$. La capa estará caracterizada por las $N\times M$ componentes de la matriz $A$ y las $N$ componentes del vector $\vec{c}$, que serán los que se entrenarán. Por tanto, la capa tendrá una cantidad de $N(M+1)$ parámetros. Para reducir el número de parámetros podemos representar la matriz $A$ como una forma MPO aproximada con una cierta dimensión de enlace $b$ de forma que tengamos un menor número de parámetros que en la versión densa original. Podemos hacer lo mismo con el vector $\vec{c}$ obteniendo su forma MPS aproximada. De esta manera, hemos comprimido la capa del modelo. Esto es favorable tanto para poder ahorrar espacio en memoria como para entrenar más rápido el modelo y evitar problemas como el overfitting, dependiendo de la representación utilizada y el modelo original.

Debido a que la aproximación de la matriz $A$ y la aproximación del vector $\vec{c}$ pueden hacerse arbitrariamente cercanos a los originales aumentando las dimensiones de enlace $b$, siempre podemos partir del modelo original e irlo comprimiendo hasta saturar un error preestablecido como tolerable. Sin embargo, también podemos comprimir los elementos del modelo y ejecutarlo para ver su precisión con respecto al modelo original, aumentando la compresión hasta que alcancemos dicho error máximo.

Otra posibilidad más sería usar una versión comprimida del modelo sin entrenar, con una inicialización controlada como en el original, y entrenar directamente el modelo comprimido. Esto presenta la ventaja de que no se pierde precisión por aproximar nuestro modelo, sino que se entrena un modelo ya preaproximado. También podemos inicializar el modelo comprimiendo uno ya entrenado y reentrenándolo.

Hay muchos esquemas de compresión además de los MPS y MPO, como los PEPS \cite{PEPS,PEPS2} o el MERA \cite{MERA}. Estos permiten realizar eficientemente ciertas operaciones en sus representaciones comprimidas y aplicar diversos métodos de optimización, como el Density Matrix Renormalization Group (DMRG) \cite{MERA,DMRG}.

\subsection{Aplicación de operaciones de gran dimensionalidad}\label{ssec: gran dim}
Un problema que podemos tener en casos industriales reales es la necesidad de aplicar operaciones de gran dimensionalidad, como crear ciertos kerneles para métodos como el Support Vector Machine \cite{SVM}. Podemos poner como ejemplo el querer aplicar un kernel en producto tensorial de las componentes de un vector de entrada $\vec{x}$ de $N$ componentes \cite{Anomaly}. Este kernel es
\begin{equation}\label{eq: kernel producto}
    \phi(\vec{x})_{i_1, i_2, \dots, i_{N}} = \phi_1(x_1)_{i_1} \phi_2(x_2)_{i_2} \dots \phi_{N}(x_{N})_{i_{N}},
\end{equation}
donde cada kernel $\phi_j(\cdot)$ afecte a la componente $j$-ésima del vector $\vec{x}$. En este caso, podemos considerar un kernel
\begin{equation}
    \phi_j(x_j)=(x_j, 1), 
\end{equation}
de forma que el kernel global $\phi(\cdot)$ nos dará todos los posibles productos entre las componentes del vector de entrada, como un tensor de $2^N$ componentes que nos sirva para expresar correlaciones en el vector. Como vemos, de manera estándar esto requeriría una cantidad de memoria y tiempo exponencial en el tamaño del vector de entrada. Además, si queremos aplicar una matriz a dicho kernel necesitaremos una matriz $A$ de dimensiones $M\times2^{N}$, lo cual escala prohibitivamente.

\begin{figure}[htb] 
\begin{center} 
  \includegraphics[width=7.5cm]{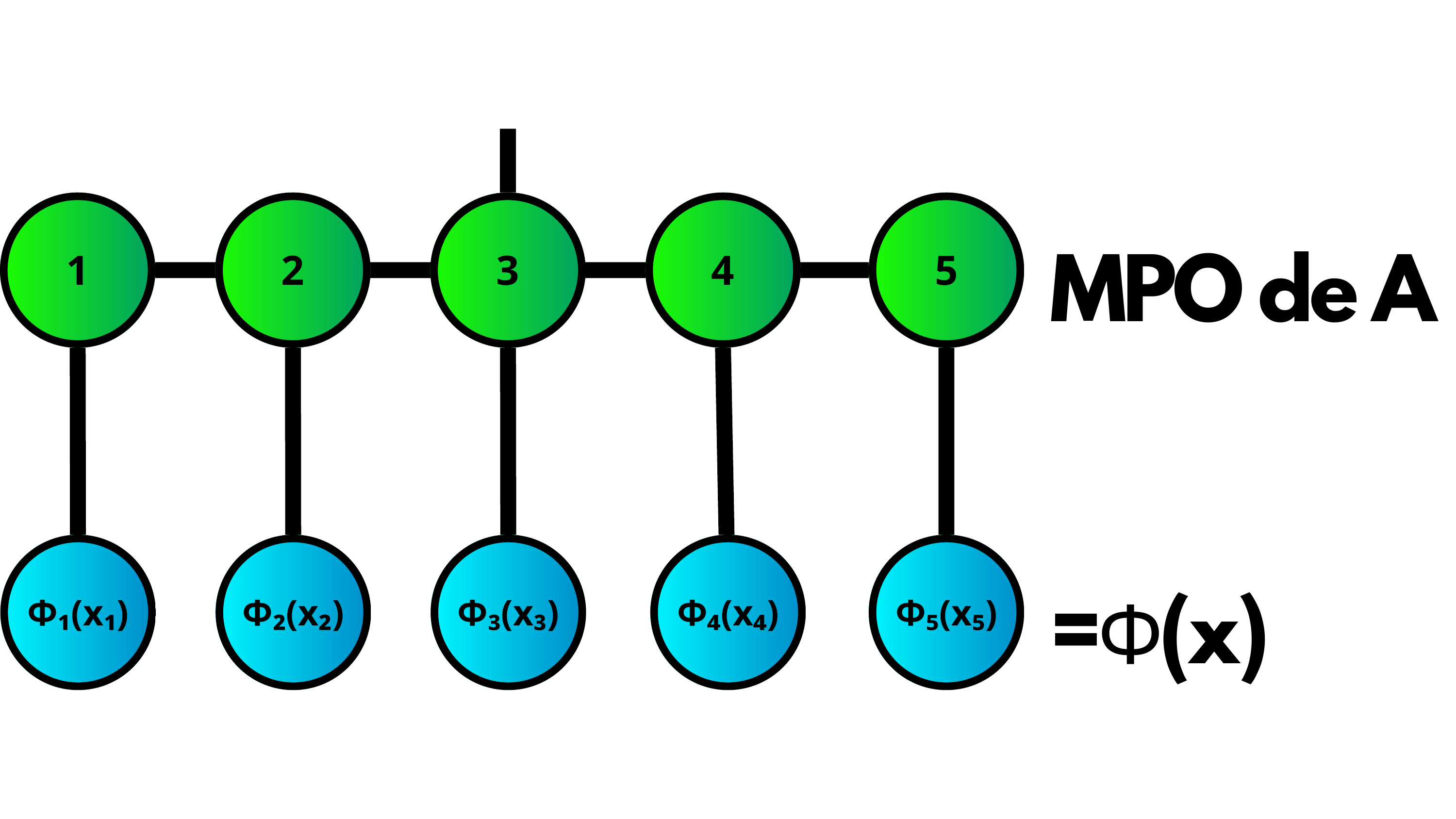}
\end{center} 
\caption{Kernel de la Ec. \ref{eq: kernel producto} y capa MPO para un vector de entrada de $N=5$ componentes.} 
\label{fig: Kernel_y_MPO} 
\end{figure}

Sin embargo, aprovechando las propiedades de las tensor networks podemos utilizar una representación MPO para dicha matriz, de la forma representada en la Fig. \ref{fig: Kernel_y_MPO}, donde tan solo es necesario contraer $2N$ tensores más pequeños.

\begin{figure}[htb] 
\begin{center} 
  \includegraphics[width=7.5cm]{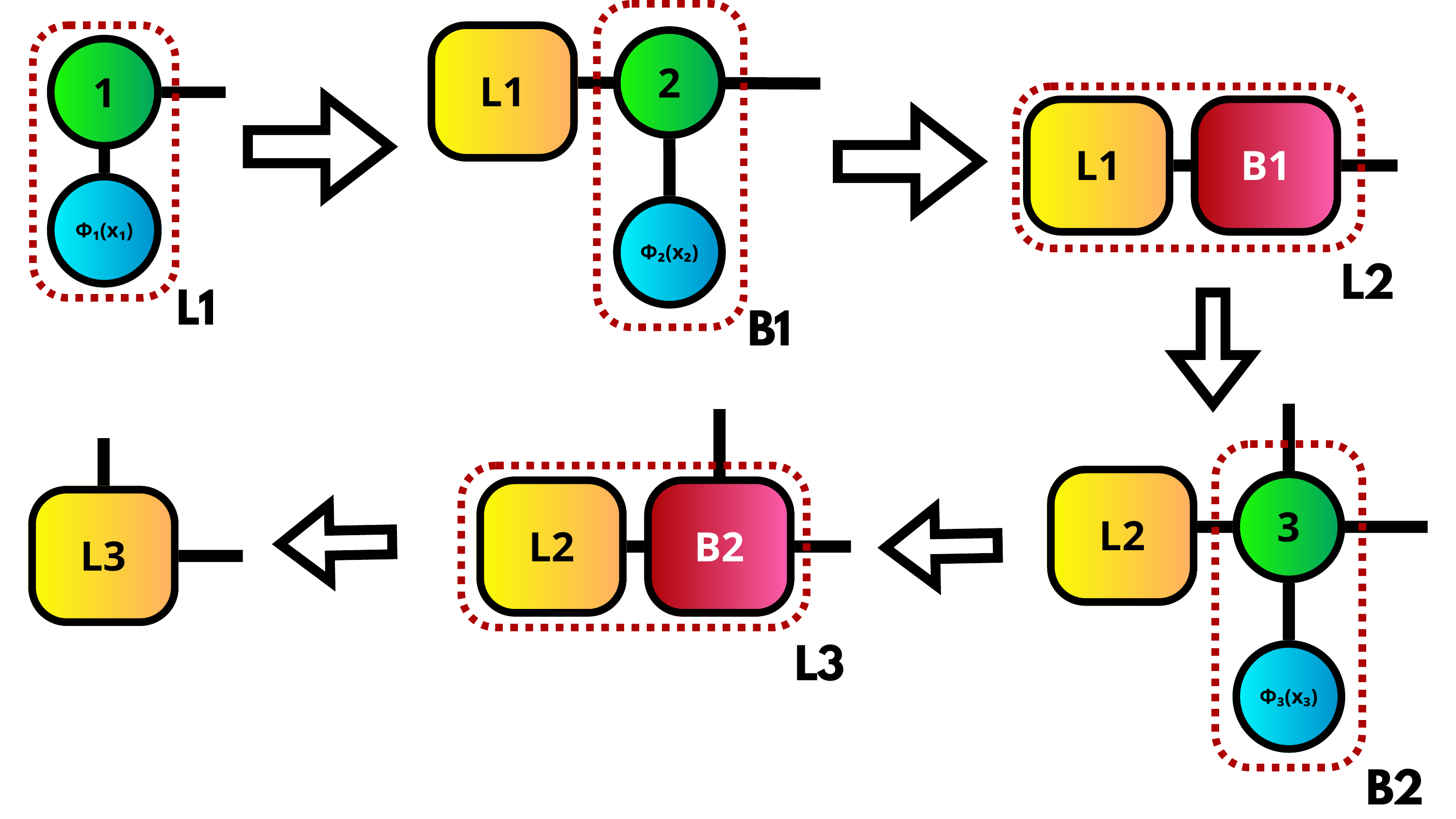}
\end{center} 
\caption{Contracción de los tres primeros nodos de la capa MPO con los tres primeros nodos del kernel. 1) Contraemos un nodo del kernel con su correspondiente de la capa MPO. 2) Contraemos el siguiente nodo kernel con su nodo de la capa. 3) Contraemos los nuevos dos nodos obtenidos. 4) Repetimos el proceso de contracción para la siguiente pareja de nodos. 5) Contraemos los nodos actuales, manteniendo el índice físico.} 
\label{fig: Contracción} 
\end{figure} 

Con esta metodología podemos aplicar una operación en la que ponderamos los elementos del tipo $x_1x_2x_3x_4x_5$, $x_1x_2x_3x_4$, $x_1x_2x_3$, $x_1x_2$ y $x_1$, de manera eficiente y con una cantidad de elementos reducida.

\subsection{Compresión de datos y eliminación de ruido}
En el campo del big data existe la necesidad de trabajar con grandes volúmenes de datos, lo cual puede ser muy costoso o irrealizable para ciertos sistemas. La propiedad de compresión de tensores que hemos visto en las representaciones como la MPS permiten comprimir \cite{Big_Data} y manejar \cite{Big_Data2} grandes cantidades de datos. Para ello solo necesitamos convertir los datos en un tensor y comprimirlo. Además, si nos permitimos reducir la precisión de sus elementos o solo estamos interesados en sus características globales, podemos usar representaciones aproximadas del tensor para ahorrar espacio en memoria.

\subsection{Optimización combinatoria}
La resolución de la mayoría de problemas de optimización combinatoria requiere tiempos de ejecución exponencialmente crecientes con el tamaño del problema. Dadas las propiedades de la computación cuántica, muchos pueden ser abordados a través de la misma de manera eficiente. Una aproximación similar a la cuántica reside en el uso de tensor networks para optimización combinatoria. Existen diferentes tipos de resolvedores de optimización combinatoria.

El primero es el desarrollado en \cite{TTOpt}, funcionando en una caja negra. Este algoritmo parte de un tensor de costes, $T$, cuyos subíndices son las componentes del vector solución. Este tensor tendrá el coste de todos los posibles estados, aunque no hace falta conocer el tensor de forma explícita: basta con tener una función que nos dé el coste en función de sus índices. Con este tensor, realizaremos una aproximación de su \textit{max volume submatrix}, la submatriz que contiene sus mayores valores singulares en producto, mediante una representación MPS. De esta representación obtendremos, con alta probabilidad, el elemento de menor coste. 

El problema de este algoritmo es que cuando trabajamos con problemas con muchas configuraciones no válidas, tiende a fallar por ver que casi todos o todos los elementos son iguales, ya que va muestreando el tensor. Por ello, necesitamos una codificación, por complicada que sea, en la cual reduzcamos el espacio de configuraciones no válidas, a fin de que el algoritmo vea un cierto mapa funcional. 

La segunda posibilidad se basa en el uso de Inteligencia Artificial (IA) Generativa potenciada con tensor networks \cite{GEO}.

La tercera posibilidad es el uso de la evolución en tiempo imaginario a un estado cuántico en superposición uniforme de todas las combinaciones posibles, equivalente a la aplicación de una puerta Hadamard a cada qubit de un circuito cuántico. Si desde esta superposición uniforme amortiguamos cada combinación de manera exponencial a su coste asociado, la combinación óptima será la que tenga una mayor amplitud, resultando en una mayor probabilidad de medida.

Debido a su no-unitareidad, a nivel cuántico no se puede realizar este tipo de operación, pero a nivel clásico sí \cite{Combin}. Si necesitásemos tener en cuenta todas las combinaciones posibles, necesitaríamos una cantidad inasumible de memoria, resultando en una búsqueda por fuerza bruta. Sin embargo, combinando la capacidad de compresión de información y simulación de estados cuánticos con la capacidad de aplicar operaciones no unitarias de las tensor networks, podemos crear dicho estado para diversos problemas \cite{TN_TSP,TN_QUBO,TN_Task}. Esto puede realizarse de manera exacta o aproximada, mediante recompresiones, dependiendo de cuanta memoria queramos dedicarle.

La principal limitación reside en que para los problemas NP-Hard, los más complejos, necesitaremos una cantidad exponencial de memoria y tiempo para resolverlos de manera exacta. Realizando ajustes inteligentes en la codificación y expresión del problema, la creación y contracción de la tensor network y aplicando diversas simplificaciones en las restricciones, podemos crear algoritmos aproximados eficientes a partir de los exactos, evitando un escalado exponencial.

La cuarta es la simulación de un sistema físico reresentado por una tensor network variacional, con una función de coste dada como un hamiltoniano que podamos escribir en representación MPS o MPO. En este caso realizaremos actualizaciones variacionales de la tensor network del estado, usando técnicas como el descenso de gradiente o el DMRG \cite{DMRG} para minimizar la energía del sistema y así encontrar la combinación óptima. La efectividad de este tipo de técnicas depende mucho del problema al que se apliquen y pueden quedar atascadas en mínimos locales, sobre todo para problemas con restricciones. Además, existe la limitación de que debemos poder expresar la función de coste en una representación de tensor network.

\section{Estudio de casos de uso}
En esta última sección presentaremos los principales casos de uso en los cuales podemos ver un potencial directo y significativo de las técnicas de tensor networks a contextos industriales y empresariales.
\subsection{Finanzas}
En el ámbito financiero contemporáneo, la alta dimensionalidad surge como un fenómeno desafiante. Con el rápido avance tecnológico y la disponibilidad de datos a gran escala, los mercados financieros se han vuelto intrínsecamente multidimensionales. Gracias a ello, se puede hacer un análisis profundo del mercado.

\subsubsection{Optimización de portfolios}
La optimización de portfolios consiste en la determinación de la mejor distribución de activos financieros en los cuales invertir para maximizar una función de coste, que suele ser aquella que maximiza el retorno y minimiza el riesgo. Este problema ya es complicado de manera estática, volviéndose extremadamente complejo de manera dinámica, con una cartera que deberemos optimizar sobre una serie de días de comercio. La alta dimensionalidad y las restricciones del problema hacen que en general sea un problema intratable de manera exacta.

Las tensor networks se pueden utilizar para explorar el espacio de soluciones de inversión a fin de buscar la combinación óptima. En \cite{Portfolio} se puede observar cómo se usa la evolución en tiempo imaginario sobre una representación MPS para resolver el problema de optimización. Consiguen unos resultados de sharpe ratio, una medida del exceso de rendimiento por unidad de riesgo en una inversión, superiores a los obtenidos por otros métodos para tamaños grandes, aunque con menor beneficio y mucho mayor tiempo de ejecución para problemas grandes que el resolvedor híbrido de Dwave.

\subsubsection{Predicciones interpretables}
En \cite{Prediccion} se presenta un método de redes neuronales recurrentes para realizar predicciones interpretables, aquellas en las cuales se pueda entender el razonamiento subyacente, utilizando una representación MPS de la matriz de pesos de la capa recurrente. Se puede observar que los resultados tanto de precisión como de Sharpe y retorno son superiores en la versión con tensor network que en las demás versiones comparadas.

\subsection{Medicina}
El campo de la medicina es uno de los que presenta mayores dimensionalidades, tanto para el diagnóstico y tratamiento de enfermedades, como en el diseño y producción de fármacos. Debido a ello, recientemente ha habido un gran interés en la aplicación de técnicas de Inteligencia Artificial en este campo, obteniendo grandes resultados.

\subsubsection{Descubrimiento de fármacos}
La simulación de fármacos es un desafío monumental en la ciencia contemporánea debido a la complejidad de las interacciones moleculares y la enorme cantidad de posibles configuraciones. En este contexto, la computación cuántica ofrece una clara ventaja ya que es capaz de modelar y tratar de manera más adecuada que la clásica fenómenos como la estructura electrónica de las moléculas y las interacciones entre átomos. 

De manera similar, podemos ahorrar tiempo y costes de descubrimiento de nuevos fármacos mediante la predicción de nuevas propiedades terapéuticas asociadas a los mismos. Esto se puede lograr de diversas maneras, siendo una de ellas la descomposición tensorial \cite{Drug}. Para ello creamos un tensor fármaco-gen-enfermedad, que dé cuenta de las relaciones encontradas entre estas parejas fármaco-gen, fármaco-enfermedad y gen-enfermedad. Con esto, formulamos un problema de completado de tensor. El tensor inicial tendrá solo las relaciones directas de las parejas que nosotros hemos encontrado, mientras que el resto de relaciones más complejas serán las que tengamos que determinar. Para ello, realizaremos una descomposición generalizada de tensor (GTD), para modelar parte de estas relaciones, y aplicaremos un perceptrón multicapa (MLP) para capturar el resto de relaciones.

La principal limitación de este método es que su capacidad predictiva es inferior para nuevas enfermedades, por lo que deben acoplársele otras técnicas.

\subsubsection{Análisis de imágenes médicas}
El análisis de imágenes, y sobre todo imágenes médicas, es una tarea muy complicada, ya sea por las características de dichas imágenes, por el ruido en las mismas o por la gran dimensionalidad que puede alcanzar el problema. Sin embargo, como hemos visto, precisamente las tensor networks son ideales para casos con alta dimensionalidad y ruido.

En \cite{Imagemed} se muestran métodos para clasificar imágenes médicas, utilizando exactamente el esquema que explicamos en la Ssec. \ref{ssec: gran dim}, interpretando las imágenes como un vector de entrada y un feature map \mbox{cosenoidal}, y la capa MPO como un clasificador. Así, al contraer la tensor network obtendremos un vector que nos dirá las probabilidades de que esa imagen pertenezca a un grupo u otro. Esto se realiza en combinación con otras técnicas más avanzadas de machine learning y finalmente se obtiene un nivel de compresión extremadamente alto con unos niveles de área bajo la curva de ROC (Receiver Operating Characteristic) similares o superiores a los métodos con los que se compara. Todo esto con unos requerimientos de GPU muy inferiores.

La segmentación de imágenes médicas es vital para el diagnóstico de enfermedades, la monitorización de su tratamiento, la guía para procesos quirúrgicos, el desarrollo de fármacos experimentales y la educación médica. En \cite{Image3d} se muestra un método para la segmentación de imágenes médicas con transformers tensorizados, introducidos en el paper. En este método, se realiza una versión tensorizada del self-attention module, un paso del transformer.

\subsection{Simulación de materiales cuánticos y materiales topológicos}
La simulación de materiales cuánticos y materiales topológicos es un problema extremadamente complejo, debido a que son problemas de alta dimensionalidad. Los sistemas cuánticos de muchos cuerpos requieren una memoria exponencial en el número de elementos para representar sus funciones de onda. La utilidad de estos materiales es enorme, dadas sus curiosas aplicaciones, como la electrónica de alta velocidad, células solares, sensores cuánticos, telecomunicaciones y computación cuántica.

Para lidiar con esto, se proponen ansatz concretos
\cite{Materials,Topology1,Topology2,PEPS_Methods} los cuales representan de manera precisa y eficiente los estados de mínima energía de estos sistemas. Se empieza con un estado inicial y se le realizan diversas transformaciones, como DMRG o evolución en tiempo imaginario, para que caigan a su estado de mínima energía. Una vez obtenido dicho estado, podemos obtener fácilmente propiedades del mismo aplicándole los tensores de las magnitudes a medir.

\subsection{Simulación de baterías}
La simulación de baterías es un problema de gran utilidad, pero que también requiere el estudio de sistemas cuánticos de muchos cuerpos. Para esto también podemos utilizar ansatz en tensor networks \cite{Baterias,Baterias2}, con los cuales podemos obtener propiedades de los estados de mínima energía.

\subsection{Optimización}
Los problemas de optimización en entornos industriales son muy importantes, ya que pueden requerir la optimización del orden en el que se realizan procesos, como la asignación de tiempos y tareas a máquinas en el Job Shop Scheduling Problem (JSSP). Sin embargo, debido al tamaño de problema en estos casos, resulta muy costoso resolver dichas instancias debido a la alta cantidad de combinaciones y la complejidad de las funciones de coste.

\subsubsection{Optimización de rutas}
El Traveling Salesman Problem (TSP) es un problema histórico ampliamente estudiado, en el cual se debe escoger una ruta que recorra todos los nodos de un cierto grafo una única vez y ofrezca el menor coste asociado a la misma, dado por los costes de ir de un nodo a otro. Generalizaciones y casos particulares de este problema modelizan problemas de reparto, muy importantes en la industria.

Para ello, en \cite{TN_TSP} se propone un método en TN que puede resolver casos de TSP generalizado, con diferentes variantes. Dicho método toma como variable el nodo en el que se está en cada paso de la ruta y así modelizar la función de coste como un Quadratic Unconstrained Discrete Optimization (QUDO) al vecino más próximo con restricciones de no repetición. Su mecánica básica reside en la evolución en tiempo imaginario descrita en \cite{TN_QUBO} y la conexión de un conjunto de capas que imponen las restricciones, eliminando las combinaciones no compatibles en el estado representado. Modificando dichas capas de restricción y la modelización del QUDO se puede abordar diferentes generalizaciones, como el poder repetir hasta $N$ veces un cierto nodo, que el inicio y el final puedan ser diferentes o que solo podamos pasar una vez por cada grupo de nodos. Todos esos casos aparecen estudiados en \cite{TN_TSP}.

La principal limitación de este método es el escalado exponencial que sufre con el número de nodos debido a las restricciones, pero métodos iterativos como el explorado en \cite{TN_Task} son capaces de limitar este escalado con éxito.

\subsubsection{Optimización de asignación de puestos}
Un caso particular del TSP generalizado es la asignación de puestos de trabajo a trabajadores. Supongamos un conjunto de trabajadores que ya están asignados a un conjunto de puestos y tenemos una serie de vacantes. El objetivo será redistribuir a los trabajadores al conjunto de puestos, incluyendo los ocupados y los vacantes, tales que maximicemos una cierta función de coste y solo haya 0 o 1 trabajadores por puesto. Este problema ha sido estudiado y tratado al final de \cite{TN_TSP}, utilizando la formulación de \cite{JRP}, la cual busca maximizar una combinación entre la productividad del puesto y la compatibilidad del trabajador asignado al mismo.

\subsubsection{Optimización de secuencias de fabricación}
La optimización de secuencias de fabricación es un problema de gran utilidad industrial. Tenemos un conjunto de productos a fabricar de manera secuencial, tales que existe un tiempo adicional de fabricación al pasar de fabricar una clase de producto a fabricar otra. Este tiempo puede ser un tiempo de cambio de piezas de una máquina, de configuración o de materiales. Por ello queremos obtener la secuencia que minimice este tiempo adicional global.

Este caso aparece estudiado en \cite{TN_TSP}, ya que se puede ver como un TSP en el cual podemos estar en cada nodo tantas veces como productos pendientes haya de la clase asociada a ese nodo en el conjunto inicial. Con esta perspectiva, un nodo representa una clase en vez de un producto para optimizar las capas de restricción.

\subsubsection{Optimización de asignación de tareas a máquinas}
La asignación óptima de tareas a máquinas se puede modelar como la elección de ejecución de tareas en un conjunto de máquinas tales que hay una restricción entre las tareas que realizan unas y otras. Un ejemplo de restricción sería el siguiente: `si la máquina 1 hace un cortado, la máquina 2 no puede hacer un pelado'. Estas restricciones pueden ser extraídas de un registro histórico o del conocimiento del negocio del gestor habitual de la máquina. Además, cada tarea tendrá un tiempo de ejecución propio en su máquina correspondiente.

Este problema es abordado en \cite{TN_Task}, donde se combinan un método de TN con evolución en tiempo imaginario y capas de restricciones como las de \cite{TN_TSP} con un algoritmo genético y un método iterativo para reducir el escalado en tiempo y memoria de dicho algoritmo, aprovechando propiedades de este tipo de problemas. La principal limitación del método es que, aunque tiene una alta tasa de éxito, para determinados casos no puede obtener soluciones satisfactorias para un límite de memoria dado.

\subsection{Big Data}
Tal como indica su nombre, en el campo del big data se manejan grandes volúmenes de datos, por lo que necesitamos sistemas específicos para tratar con ellos.

En \cite{Big_Data} se presentan diferentes métodos para tratar con problemas de big data mediante descomposiciones de tensor networks, además de otros temas. Por otro lado, en \cite{Big_Data2} se presenta un método de tensor networks para compartición de secretos (o `secret sharing') en big data, basado en la descomposición en un MPS o una descomposición de Tucker \cite{Tucker} del tensor de datos original mediante TSVD consecutivas. Entre cada TSVD se realizará una perturbación del tensor, mediante una multiplicación del nuevo nodo obtenido de la SVD por una matriz $\Delta$, y al tensor que deberemos descomponer en el paso siguiente por $\Delta^{-1}$. Esto consigue que el tensor global sea el mismo, pero los nodos de la representación sean diferentes. Finalmente se aleatoriza el primer nodo de la representación. De esta manera, se pueden guardar partes del dataset de manera distribuida manteniendo la seguridad de los mismos, ya que nadie podrá obtener ninguna información de dichos datos hasta tener todos los nodos de la representación. A su misma vez, presenta una manera de realizar computación distribuida para big data.

\subsection{Clasificación}
Como hemos visto en el caso de la clasificación de imágenes médicas, usando métodos como el presentado se pueden clasificar diferentes tipos de imágenes o datos generales.
Además, existen métodos más generales. Por un lado, la inspiración de circuitos cuánticos en métodos de tensor networks para hacer circuitos de clasificación \cite{ImageTN}. En este caso se abordan los árboles jerárquicos.

Por otro lado, tenemos el uso de PEPS para la clasificación de imágenes, donde se consiguen unos resultados similares a los de GoogLeNet, VGG-16 y AlexNet  \cite{ImageTN2} mediante sistemas simples de tensor networks acoplados.

\subsection{Inteligencia Artifical}
Como vimos en Ssec. \ref{ssec: mach}, diversos modelos dedicados a inteligencia artificial pueden ser comprimidos mediante representaciones de tensor networks. Combinando eso con lo mostrado en Ssec. \ref{ssec: gran dim}, podemos tratar problemas mucho más grandes y complejos. Además, podemos añadir diversas técnicas de machine learning cuántico \cite{QML} para diseñar nuevos modelos con las características que necesitemos.

Por otro lado, la creación de modelos más compactos que puedan insertarse y entrenarse directamente en dispositivos más modestos, como dispositivos móviles, puede permitir la experimentación con más modelos de manera simultánea y nuevas aplicaciones.

También podemos realizar técnicas de machine learning que preserven la privacidad mediante técnicas de tensor networks \cite{Privacy}. Esto tiene una vital importancia cuando se trabaja con datos sensibles, como los registros médicos.

\subsection{Ciberseguridad}
La ciberseguridad es un campo muy complejo, en el cual se necesitan medidas extraordinarias para proteger la privacidad de los usuarios y detectar posibles ataques \cite{CiberTN}. Aquí podemos englobar las técnicas ya vistas anteriormente aplicadas a este contexto.

\subsection{Detección de anomalías}
La detección de anomalías es un problema consistente en identificar casos inusuales en un conjunto de datos. Este problema es muy útil a la hora de detectar ataques informáticos. Debido a que los datos anómalos pueden presentar muchas estructuras, teniendo un gran espacio de posibilidades, se requiere modelizar de una manera muy efectiva los datos normales de forma que puedan ser identificados con respecto a los anómalos. Esto suele requerir analizar datos de gran dimensionalidad o aplicar operaciones muy complejas a los mismos.

En \cite{Anomaly} se analiza un método supervisado precisamente para poder lidiar con este problema. Aplicando un ansatz producto como el que explicamos en la Ssec. \ref{ssec: gran dim}, y una operación en MPO variacional que entrenaremos, podemos proyectar los datos normales a una hiperesfera. De esta manera, los puntos normales quedarán sobre la superficie de la esfera, mientras que los anómalos irán al centro o al exterior de la misma. Además, el proceso puede realizarse de manera muy eficiente debido a la estructura de la tensor network a contraer, escalando notablemente bien con el tamaño del problema.

Ha sido probado con diferentes datasets contra diferentes algoritmos, dando buenos resultados \cite{Anomaly}. Su principal limitación es que es un algoritmo supervisado, en el cual requerimos un etiquetado previo bien realizado para que funcione debidamente. Aun así, puede ser extendido para ser parte de un algoritmo no supervisado añadiendo un preetiquetado automático.

\section{Conclusiones}
Hemos visto que las tensor networks son un conjunto de técnicas con gran aplicabilidad al contexto industrial de manera inmediata, con ventajas probadas y una extensa literatura al respecto. Hemos explicado las principales características que las hacen especialmente atractivas para problemas de alta dimensionalidad y para el machine learning. También hemos destacado diversos casos aplicados y algunas de las técnicas que se han utilizado en el estado del arte para abordarlos, además de sus ventajas y desventajas.

\section*{Agradecimientos}
La investigación presentada en este artículo ha sido financiada como parte del proyecto Q4Real (Quantum Computing for Real Industries), HAZITEK 2022, no. ZE-2022/00033.

\nocite{*}
\bibliographystyle{Jornadas}
\bibliography{biblio}

\end{document}